\documentstyle[multicol,aps,psfig]{revtex} 
 
\begin{document} 
 
\draft 
\preprint{LBL-45309} 
 
\title{Gluon Shadowing and Hadron Production at RHIC} 
 
\author{Shi-yuan Li$^a$ and Xin-Nian Wang$^{b,a}$} 
\address{$^a$Department of Physics, Shandong University, Jinan, P.R.C., 250100\\
         $^b$ Nuclear Science Division, Mailstop 70-319, 
         Lawrence Berkeley Laboratory, Berkeley, California 94720.} 
\date{October 3,2001} 

\maketitle 
\begin{abstract} 
Hadron multiplicity in the central rapidity region of high-energy heavy-ion 
collisions is investigated within a two-component mini-jet model which 
consists of soft and semi-hard particle production. The hard 
contribution from mini-jets is reevaluated using the latest parameterization 
of parton distributions and nuclear shadowing. The energy dependence of 
the experimental data from RHIC requires a strong nuclear shadowing of 
the gluon distribution in this model. The centrality dependence of the 
hadron multiplicity at $\sqrt{s}=130$ GeV is reproduced well with the 
impact-parameter dependent parton shadowing. However, energy variation 
of the centrality dependence is needed to distinguish different 
particle production mechanisms such as the parton saturation model.
\end{abstract} 
 
 
\begin{multicols}{2}  

Formation of quark-gluon plasma (QGP) in high-energy heavy-ion
collisions hinges crucially on the initial condition that is
reached in the earliest stage of the violent nuclear interaction.
Though many proposed signals can provide more direct measurements of
the initial parton density, they must compliment results inferred
indirectly from the measurement of final hadron multiplicity using
either simple scenarios such as the Bjorken model \cite{bj} or
other dynamic models. Therefore, global observables
such as the rapidity density of hadron multiplicity can provide
an important link of a puzzle that can eventually lead one to a
more complete picture of the dynamics of heavy-ion collisions 
and formation of QGP. Furthermore, the study of energy and centrality
dependence of central rapidity density \cite{wg01} can also provide 
important constraints on models of initial entropy production and
shed lights on the initial parton distributions in nuclei.
For example, the available RHIC experimental data 
\cite{phob1,phob2,phenix,star,brahms} can already rule out the
simple two-component model without nuclear modification of the
parton distributions in nuclei \cite{wg01}. In this paper,
we will study within the two-component model how the RHIC
data constrain the unknown nuclear shadowing of the gluon 
distribution in nuclei and how to further distinguish such a 
conventional parton production mechanism from
other novel physics such as parton saturation \cite{kn,kl}. 

Mini-jet production in a two-component model has long been proposed 
to explain the energy dependence of total cross section 
\cite{gaisser,pancheri} and particle production \cite{chen,wang91}
in high-energy hadron collisions. It has also been 
proposed \cite{eskola,bm} and incorporated 
in the HIJING model \cite{hijing,wang97} to describe initial 
parton production in high-energy heavy-ion collisions. In this
simple two-component model, one assumes that events of 
nucleon-nucleon collisions at high energy can be divided into
those with and without hard or semi-hard processes of jet production.
The soft and hard processes are separated by a cut-off scale $p_0$.
While the cross section of soft interaction $\sigma_{\rm soft}$ is 
considered nonperturbative and thus noncalculable, the jet production
cross section $\sigma_{\rm jet}$ is assumed to be given by 
perturbative QCD (pQCD) for transverse momentum transfer $p_T>p_0$.
The two parameters, $\sigma_{\rm soft}$ and $p_0$, are determined
phenomenologically by fitting the experimental data of total 
$p+p(\bar{p})$ cross sections within the two-component
model \cite{gaisser,pancheri,chen,wang91,hijing,wang97}.

The cut-off scale $p_0$, separating nonperturbative and pQCD
components, could in principle depend on both energy
and nuclear size. Using Duke-Owens parameterization \cite{do}
of parton distributions in nucleons, it was found in the 
HIJING \cite{hijing} model that an energy independent cut-off
scale $p_0=2$ GeV/$c$ is sufficient to reproduce the experimental data
on total cross sections and the hadron multiplicity in $p+p(\bar{p})$
collisions, assuming that the soft cross section $\sigma_{\rm soft}$ 
is also constant. Since then, analysis of recent experimental data 
from deep-inelastic scattering (DIS) of lepton and nucleon at HERA 
indicated \cite{grv} that gluon distribution inside a nucleon is 
much larger than the DO parameterization at small $x$. Many new
parameterizations of the parton distributions have become available. 
Using the Gluck-Reya-Vogt (GRV) parameterization \cite{grv} of parton
distributions and following the same procedure as in the original
HIJING \cite{hijing}, we find that one can no longer fit the 
experimental $p+p(\bar{p})$ data using a constant cut-off scale $p_0$
within the two-component model. One has
to assume an energy dependent cut-off scale $p_0(\sqrt{s})$.
Because of the rapid increase of gluon distribution at small $x$,
we find that the cut-off $p_0(\sqrt{s})$ has to increase slightly
with energy in order to fit the experimental data.

\begin{figure} 
\centerline{\psfig{figure=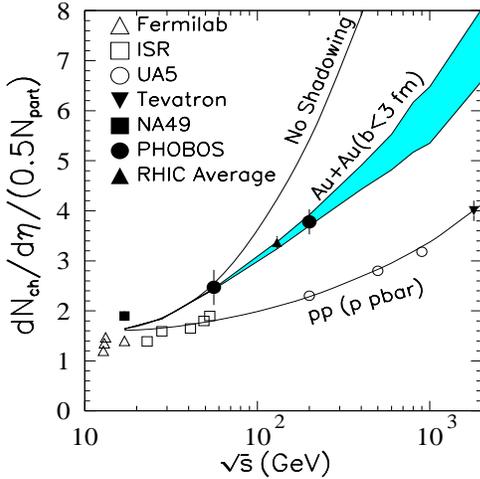,width=2.5in,height=2.5in}}
\caption{
Charged particle rapidity density {\em per  participating 
nucleon pair} versus the c.m. energy. The RHIC
data\protect\cite{phob1,phob2} (filled circle and up-triangle) 
for the  6\% most central Au+Au are  
compared to $pp$ and $p\bar{p}$ data (open symbols)  
\protect\cite{ppbar,ppisr,ppfermilab} 
and the NA49 $Pb+Pb$(central 5\%) data \protect\cite{na49} (filled square).  
The two-component mini-jet model with and without shadowing is
also shown. The shaded area for central $Au+Au$ collisions corresponds
to the range of gluon shadowing parameter 
$s_g=0.24$--0.28 [Eq.~(\protect\ref{eq:shg})]. }
\label{fig1} 
\end{figure}

Shown in Fig.~\ref{fig1} is the calculated central rapidity density,
\begin{equation}
\frac{dN_{\rm ch}}{d\eta}=\langle n\rangle_{\rm s}
+\langle n\rangle_{\rm h} 
\frac{\sigma_{\rm jet}(s)}{\sigma_{\rm in}(s)},
\end{equation}
for $p+p(\bar{p})$ collisions as a function of energy $\sqrt{s}$,
where $\langle n\rangle_{\rm s}=1.6$ and $\langle n\rangle_{\rm h}=2.2$
represent particle production from soft interaction and jet
hadronization, respectively. The jet cross section in lowest order
of pQCD is given by
\begin{equation}
\sigma_{\rm jet}=K \int_{p_0^2}^{s/4} dp^2_T dy_1 dy_2
\frac{1}{2}\sum_{a,b,c,d} x_1 x_2 f_a(x_1)f_b(x_2) 
\frac{d\sigma_{ab\rightarrow cd}}{d\hat{t}},
\label{eq:jet}
\end{equation}
with the GRV parameterization \cite{grv} of parton distributions
$f_a(x)$, a $K$-factor of 2 and an energy-dependent cut-off scale
\begin{eqnarray}
p_0(\sqrt{s})&=&3.91-3.34\log(\log\sqrt{s})+0.98\log^2(\log\sqrt{s})
\nonumber \\
&+&0.23\log^3(\log\sqrt{s}). 
\label{eq:cut}
\end{eqnarray}
Assuming eikonalization of hard and soft processes, the
total inelastic $p+p(\bar p)$ cross section in this two-component
model is \cite{hijing},
\begin{equation}
\sigma_{\rm in}=\int d^2b \left[1-
e^{-(\sigma_{\rm jet}+\sigma_{\rm soft})T_{NN}(b)}\right],
\end{equation}
where $T_{NN}(b)$ is the nucleon-nucleon overlap function and
$\sigma_{\rm soft}=57$ mb represents the inclusive soft cross 
section. Notice that the energy-dependence of $p_0(\sqrt{s})$ is 
quite weak, ranging from $p_0=1.7$ GeV/$c$ at $\sqrt{s}=20$ GeV to
3.5 GeV/$c$ at $\sqrt{s}=5$ TeV.

To extrapolate the two-component model to nuclear collisions,
one assumes that multiple mini-jet production is incoherent and
thus is proportional to the number of binary collisions 
$N_{\rm binary}$. The soft interaction is however coherent and 
proportional to the number of participant nucleons $N_{\rm part}$
according to the wounded nucleon model \cite{bialas}. Assuming no
final state effects on multiplicity from jet hadronization, the 
rapidity density of hadron multiplicity in heavy-ion collisions is then,
\begin{equation}
\frac{dN_{ch}}{d\eta}=\frac{1}{2}\langle N_{\rm part}\rangle 
\langle n\rangle_{s} 
+ \langle n\rangle_{h}\langle N_{\rm binary}\rangle 
\frac{\sigma_{\rm jet}^{AA}(s)}{\sigma_{\rm in}},
\label{eq:nch} 
\end{equation}
where $\sigma_{\rm jet}^{AA}(s)$ is the averaged inclusive jet cross
section per $NN$ in $AA$ collisions. The average number of
participant nucleons and number of binary collisions for given
impact-parameters can be estimated using HIJING Monte Carlo simulation.
If one assumes that the jet production 
cross section $\sigma_{\rm jet}^{AA}(s)$ is the same 
as in $p+p$ collisions, the resultant energy dependence of the 
multiplicity density in central nuclear collisions is much stronger 
than the RHIC data as shown in Fig.~\ref{fig1}. Therefore, one
has to consider nuclear effects of jet production in heavy-ion
collisions.

In high-energy nuclear collisions, multiple mini-jet production can 
occur within the same transverse area. If there are more than one pair of 
mini-jet production within the transverse area given by the jet's intrinsic 
size $\pi/p_0^2$, jet production within this area might not be 
independent any more \cite{hijing}. If such a criteria is used for
independent jet production within one unit of rapidity, 
one can then obtain a cut-off scale $p_0$ in a so-called 
final state saturation model\cite{ekrt},
\begin{equation}
p_0\approx 0.187\; A^{0.136} (\sqrt{s})^{0.205}
\end{equation}
that also depends on nuclear size for central heavy-ion collisions. 
This cut-off scale, though increasing with nuclear size, ranges 
from 0.7 GeV/$c$ at $\sqrt{s}=20$ GeV to 2.2 GeV/$c$ 
at $\sqrt{s}=5$ TeV for central $Au+Au$ collisions, which is much smaller 
than what we have obtained in Eq.~(\ref{eq:cut}) by fitting $p+p(\bar{p})$ 
data. Therefore, if we apply the two-component model to heavy-ion 
collisions with the same cut-off scale in Eq.~(\ref{eq:cut}) as 
determined in $p+p(\bar{p})$ collisions, the criteria for independent 
jet production will never be violated. Instead, we will assume 
the cut-off scale to be independent of nuclear size in this paper.

In principle, jets produced in the early stage of heavy-ion collisions
will also suffer final state interaction and induced gluon 
bremsstrahlung. For an energetic jet, this will lead to induced
parton energy loss \cite{gw94,bdms} and the suppression of large
transverse momentum hadrons \cite{wg92}. Such a jet quenching effect
could also lead to increased total hadron multiplicity \cite{wg92}
due to the soft gluons from the bremsstrahlung. However, a recent
study \cite{ww01} of parton energy loss in a thermal environment 
found that the effective energy loss is significantly reduced for 
less energetic partons due to detailed balance by thermal absorption.
Thus, only large energy jets lose significant energy via gluon 
bremsstrahlung. Since the production rates of these large energy
jets are very small at the RHIC energy, their contributions to
the total hadron multiplicity via jet quenching should also be small.
Similarly we also assume that parton thermalization during the
early stage contributes little to the final hadron multiplicity.

\begin{figure} 
\centerline{\psfig{figure=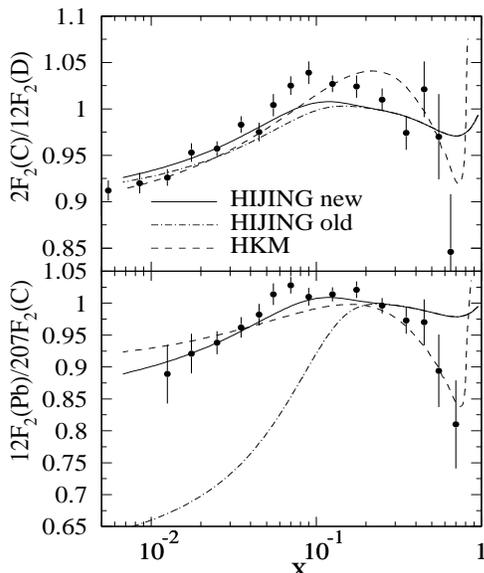,width=2.5in,height=3.0in}}
\caption{
Ratio of nuclear structure functions as measured in DIS. Solid
lines are the new HIJING parameterization [Eq.~\protect\ref{eq:shq}], 
dashed lines are the HKM parameterization \protect\cite{hkm}
and dot-dashed lines are the old HIJING parameterization 
\protect\cite{hijing}. The data are from Ref.~\protect\cite{nmc}.}
\label{fig2} 
\end{figure}

One important nuclear effect we have to consider in our two-component 
model is the nuclear shadowing of parton distributions or the depletion
of effective parton distributions in nuclei at small $x$. Such a nuclear
shadowing effect in jet production can be taken into account by assuming
modified parton distributions in nuclei,
\begin{equation}
f_a^A(x,Q^2)=AR_a^A(x,Q^2)f_a^N(x,Q^2).
\end{equation}
Using the experimental data from DIS off nuclear targets and unmodified
DGLAP evolution equations, one can
parameterize $R_a^A(x,Q^2)$ for different partons and nuclei \cite{eks,hkm}.
Recent new data \cite{nmc} however indicate that the simple
parameterization for nuclear shadowing used in HIJING \cite{hijing} 
is too strong for heavy nuclei. In this paper, we will use the following
new parameterization,
\begin{eqnarray}
R_q^A(x)&=&1.0+1.19\log^{1/6}\!\!\!A\;(x^3-1.2x^2+0.21x) \nonumber \\
        &-&s_q\;(A^{1/3}-1)^{0.6}(1-3.5\sqrt{x})\exp(-x^2/0.01)
\label{eq:shq}
\end{eqnarray}
with $s_q=0.1$ for all quark distributions as shown in Fig.~\ref{fig2}
(solid lines) in comparison with the experimental data \cite{nmc}. 
Also shown in the figure are parameterizations (dashed lines)
by Hirai, Kumano and Miyama (HKM) \cite{hkm} and the old HIJING 
parameterization \cite{hijing}. The shadowing in the old HIJING 
parameterization (dot-dashed lines) is apparently too strong for heavy 
nuclei. The HKM parameterizations also take into account constraints by
momentum sum rules, as similarly in the original parameterizations 
by Eskola,Kolhinen and Salgado (EKS) \cite{eks}. For the
purpose of this paper, one can neglect the scale dependence
of the shadowing.

\begin{figure} 
\centerline{\psfig{figure=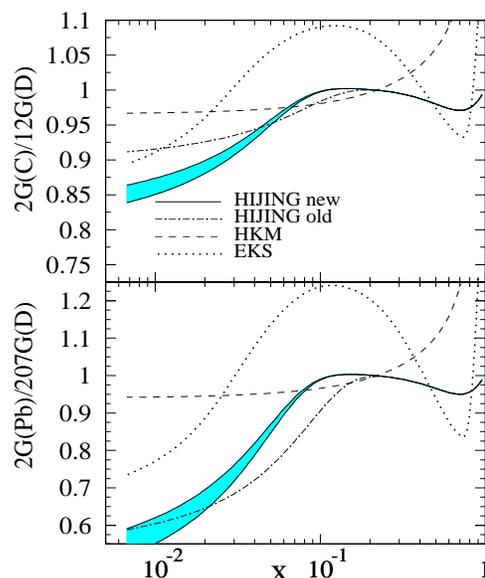,width=2.5in,height=3.0in}}
\caption{Ratios of gluon distributions in different nuclei given
by the new HIJING [Eq.~(\protect\ref{eq:shg})] (solid line, the shaded
area corresponds to $s_g=0.24$--0.28), 
old HIJING \protect\cite{hijing} (dot-dashed), 
HKM \protect\cite{hkm} (dashed) and EKS \protect\cite{eks} (dotted)
parameterization. }
\label{fig3} 
\end{figure}

The nuclear shadowing for gluons is somewhat constrained by the momentum
sum rules in the HKM parameterization. However, the constraint 
is not very strong, leaving a lot of room for large variation of
gluon shadowing. Shown in Fig.~\ref{fig3} are
the shadowing factors for gluon distribution from EKS and HKM
parameterizations. They both have strong anti-shadowing around
$x\sim 0.1$. The stronger anti-shadowing in EKS parameterization
is due to additional constraints by the $Q^2$ dependence of 
$F_2(Sn)/F_2(C)$, assuming the same unmodified DGLAP evolution
equation for parton distributions of a nucleon. Since gluon-gluon
scattering dominate the jet production cross section, such a strong gluon
anti-shadowing leads to larger jet cross section and thus larger 
hadron multiplicity than in the case of no shadowing at the RHIC 
energies. Such a scenario within the two-component model is clearly 
inconsistent with the experimental data.
We therefore propose the following parameterization for gluon shadowing,
\begin{eqnarray}
R_g^A(x)&=&1.0+1.19\log^{1/6}\!\!\!A\;(x^3-1.2x^2+0.21x) \nonumber \\
        &-&s_g\;(A^{1/3}-1)^{0.6}(1-1.5x^{0.35})\exp(-x^2/0.004),
\label{eq:shg}
\end{eqnarray}
with $s_g=0.24$--0.28. This is shown in Fig.~\ref{fig3} as the solid lines.
The hadron multiplicity density in the two-component model using
the above gluon shadowing is shown in Fig.~\ref{fig1}. The shaded area
corresponds to the variation of $s_g=0.24$--0.28. The RHIC data thus 
indicate that such a strong gluon shadowing is required within the
two-component model. If one assumes the same gluon shadowing as the
quarks in Eq.~(\ref{eq:shq}), the resultant $dN/d\eta$ is only slightly
smaller than the one without shadowing. Such a contraint on gluon
shadowing is indirect and model dependent. It is important to study
directly the gluon shadowing in other processes in $AA$ or $pA$ collisions.

\begin{figure} 
\centerline{\psfig{figure=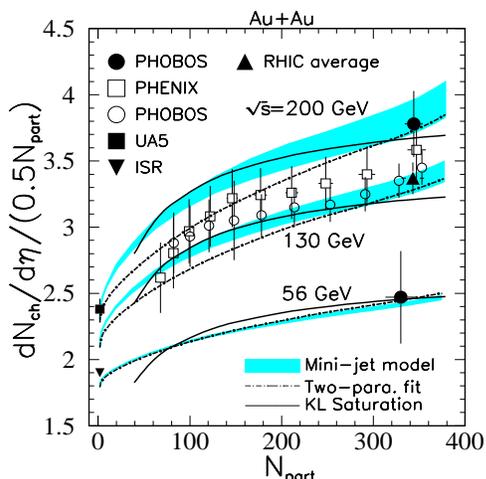,width=2.5in,height=2.5in}}
\caption{
The charged hadron central rapidity density per participant nucleon 
pair as a function of the averaged number of participants from
the two-component model (shaded lines), two-parameter 
fit [Eq.~(\protect\ref{eq:fit})] (dot-dashed lines) and parton
saturation model \protect\cite{kl} as compared to experimental
data \protect\cite{phob1,phenix,ppbar,ppisr}.}
\label{fig4} 
\end{figure}

To take into account the impact-parameter dependence of the shadowing,
we simply replace the shadowing parameters $s_a$ in Eqs.~(\ref{eq:shq}) 
and (\ref{eq:shg}) by
\begin{equation}
s_a(b)=s_a\frac{5}{3}(1-b^2/R_A^2),
\end{equation}
where $R_A=1.12 A^{1/3}$ is the nuclear size. With this impact-parameter
dependence, the calculated jet cross section 
$\sigma_{jet}^{AA}(s)/\sigma_{in}$ will also depend on the centrality
of heavy-ion collisions, decreasing from peripheral to central 
collisions. One can then calculate the centrality dependence of the hadron
multiplicity density. The results are shown in Fig.~\ref{fig4}
as a function of $N_{\rm part}$ for $Au+Au$ collisions at $\sqrt{s}=56$,
130 and 200 GeV. The shaded areas again correspond to the variation
of the gluon shadowing parameter $s_g=0.24$--0.28.  Within statistical and
systematic errors, the two-component mini-jet model with impact-parameter
dependent parton shadowing describes the PHOBOS and PHENIX 
data \cite{phob1,phenix} at $\sqrt{s}=130$ GeV well. 

To illustrate the effect of the impact-parameter dependence of the parton 
shadowing, we compare the results with a two-component parameterization,
\begin{equation}
\frac{dN_{ch}}{d\eta}=\frac{1}{2}\langle N_{\rm part}\rangle n_s 
+ \langle N_{\rm binary}\rangle n_h  ,
\label{eq:fit}
\end{equation}
shown as dot-dashed lines,
where the two parameters, $n_s$ and $n_h$, fixed at each energy
by values of $dN_{ch}^{AA}/d\eta$ in $p+p$ collisions and the most
central $Au+Au$ collisions, are assumed to be independent of
the centrality. The increase of $2dN_{ch}/d\eta/\langle N_{\rm part}\rangle$ 
with $\langle N_{\rm part}\rangle$ is driven only by the centrality
dependence of $\langle N_{\rm binary}\rangle/\langle N_{\rm part}\rangle$
in this two-parameter fit. Comparing to such a two-parameter fit, 
the two-component mini-jet model has a flatter centrality dependence 
at high energies because the effective jet cross section decreases
from peripheral to central collisions due to the impact-parameter 
dependence of parton shadowing. The better agreement between the
experimental data and the two-component mini-jet model 
at $\sqrt{s}=130$ GeV is another indication of strong nuclear
shadowing of the gluon distribution in mini-jet production.

Similar centrality dependencies are also 
predicted by other models \cite{ekrt,other},
in particular the initial-state parton saturation model \cite{kn,kl}. 
It is based on the nonlinear Yang-Mills field dynamics \cite{bm,mv}
assuming that nonlinear gluon interaction below a saturation scale
$Q_s^2\sim \alpha_s\, xG_A(x,Q_s^2)/\pi R_A^2$ leads to a classical
behavior of the gluonic field inside a large nucleus,
where $G_A(x,Q_s^2)$ is the gluon distribution at $x=2Q_s/\sqrt{s}$.
Assuming particle production in high-energy heavy-ion collisions 
is dominated by gluon production from the classical gluon field,
one has a simple form \cite{kl} for the charged hadron rapidity
density at $\eta=0$,
\begin{eqnarray}
\frac{2}{\langle N_{\rm part}\rangle}\frac{dN_{ch}}{d\eta}&=&c
\left (\frac{s}{s_0}\right)^{\lambda/2}\left[
\log\left(\frac{Q_{0s}^2}{\Lambda_{\rm QCD}^2}\right) \right.\nonumber \\
&+&\left. \frac{\lambda}{2}\log\left(\frac{s}{s_0}\right) \right ],
\end{eqnarray}
with $c\approx 0.82$ \cite{kn}. This is shown in Fig.~\ref{fig4}
as solid lines. Here, $\Lambda_{\rm QCD}=0.2$ GeV, $\lambda=0.25$ and 
the centrality dependence of the saturation scale $Q_{0s}^2$ at
$\sqrt{s_0}=130$ GeV is taken from Ref.~\cite{kn}.

\begin{figure} 
\centerline{\psfig{figure=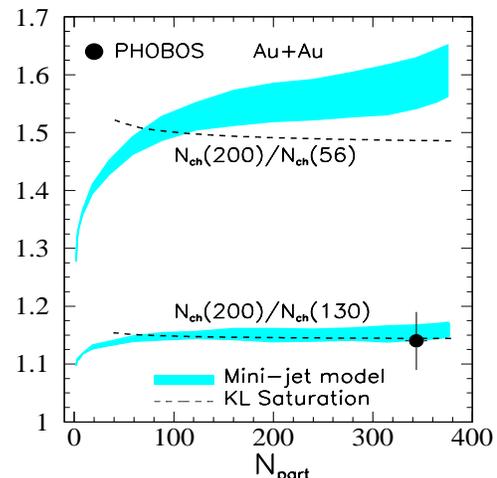,width=2.5in,height=2.5in}}
\caption{
The ratios of charged hadron multiplicity density in $Au+Au$
collisions at different energies as a function of the averaged
number of participants shown with PHOBOS data \protect\cite{phob1,phob2}.}
\label{fig5} 
\end{figure} 

Comparing the two model results in Fig.~\ref{fig4}, one notices that 
the saturation and two-component model agree with each other
in most regions of centrality except very peripheral and very central
collisions. In central collisions, results of saturation model tend
to be flatter than the two-component model. In this region, there are
still strong fluctuations in parton production in the two-component
model through the fluctuation of $N_{\rm binary}$ while $N_{\rm part}$
is limited by its maximum value of $2A$. That is why
$dN_{ch}/d\eta/\langle N_{\rm part}\rangle$ continues to increase
with $\langle N_{\rm part}\rangle$ in the central region. Such a
fluctuation is not currently taken into account in the saturation
model calculation. More accurate measurements
with small errors (less than 5\%) will help to distinguish these two
different behaviors. For peripheral collisions, saturation model results
fall off more rapidly than the mini-jet results. However, the experimental
errors are very big in this region because of large uncertainties 
related to the determination of the number of participants. Therefore,
it will be very useful to have light-ion collisions at the same energy
to map out the nuclear dependence of the hadron multiplicity in this region.
An alternative is to study the ratios of hadron multiplicity of heavy-ion
collisions at two different energies as a function of centrality. In 
this case, the errors associated with the determination of centrality
will mostly cancel. Shown in Fig.~\ref{fig5} are the ratios of hadron
multiplicity at three different energies as a function of the averaged
number of participants predicted by saturation and two-component model.
One notices that while the results from saturation model have the
same centrality dependence at all three energies the two-component
model predicts slightly different behavior at different energies,
indicating the energy dependence of the mini-jet component.
So the ratios given by the saturation model are almost independent of
centrality. On the other hand, two-component model predicts noticeable
centrality dependence of the ratios. This is especially true for
the ratio between collisions at $\sqrt{s}=200$ and 56 GeV.

It is interesting to point out that in the saturation
model that assumes a particle production mechanism dominated by coherent 
mini-jet prodctuon below the saturation scale $Q_s$, the value
of $Q_s$ determined in Ref.\cite{kn,kl} is much smaller than the 
cut-off $p_0$ in the two-component model constrained by the $p+p(\bar{p})$
data. As demonstrated in this paper, the number of mini-jet production
below such scale is still very large and should contribute to the final
hadron multiplicity. 

In summary, we have studied the energy and centrality dependence of
the central rapidity density of hadron multiplicity in heavy-ion
collisions at RHIC energies within a two-component mini-jet model.
As a consequence of the latest parameterization of parton 
distributions \cite{grv} which have a higher gluon density than
the old parameterization \cite{do} used in previous studies \cite{hijing},
the cut-off scale that separates soft and hard processes is found
to increase slightly with energy in order to fit the $p+p(\bar{p})$
data. The cut-off scale, however, is still large enough that the
independent jet production picture is still valid. With a new
parameterization of nuclear shadowing of parton distributions in
nuclei, we also found that RHIC data require a strong shadowing
of gluon distribution. Using this strong gluon shadowing with
an assumed impact-parameter dependence, the predicted centrality 
dependence of the hadron multiplicity agrees well with the recent
RHIC results. We have also compared our results with the parton
saturation model \cite{kn,kl}. We point out that in order to
differentiate the two models one needs more accurate experimental
data in both the most central and peripheral regions of centrality
or study the centrality dependence of the ratios at different
colliding energies.

\acknowledgments 
The authors thank M. Baker, K. Eskola, D. Kharzeev and P. Steinberg
for comments and discussions. This work was supported by  
the Director, Office of Energy 
Research, Office of High Energy and Nuclear Physics, 
Division of Nuclear Physics, and by the Office of Basic Energy Science, 
Division of Nuclear Science, of  
the U.S. Department of Energy 
under Contract No. DE-AC03-76SF00098 and in part by 
NSFC under project 19928511 and 10075031.

\end{multicols}  
\end{document}